\documentclass[conference]{IEEEtran}
\IEEEoverridecommandlockouts
\usepackage{cite}
\usepackage{amsmath,amssymb,amsfonts}
\usepackage{algorithmic}
\usepackage{graphicx}
\usepackage{textcomp}
\usepackage[dvipsnames]{xcolor}
\usepackage{hyperref}
\usepackage{wasysym}
\usepackage{tabularx}
\usepackage{soul}
\usepackage[normalem]{ulem}

\makeatletter
\newcommand\colorwave[1][blue]{\bgroup \markoverwith{\lower3.5\p@\hbox{\sixly \textcolor{#1}{\char58}}}\ULon}

\usepackage{fancyvrb,fvextra}
\graphicspath{{figures/}{pictures/}{images/}{../}} %

\newcommand{\etal}
{et al.}
\newcommand{\etals}
{et al.'s}

\newcommand{\ie}{{i.e.}}
\newcommand{\eg}{{e.g.}}
\newcommand{\etc}{{etc.}}
\newcommand{\cf}{{cf.}}
\newcommand{\ala}{\`a la}
\newcommand{\toolName}[1]
{#1}

\newcommand{\figref}[1]{\hyperref[#1]{Fig.~\ref*{#1}}}
\newcommand{\secref}[1]{\hyperref[#1]{Sec.~\ref*{#1}}}

{\par\kern\topsep}

\newcommand{\footnoteWithIndent}[1]
{\footnote{#1}}

\definecolor{linkColor}{HTML}{257E98}
\usepackage{soul}
\setuldepth{Berlin}
\newcommand\asLink[2]{\textcolor{linkColor}{\href{#1}{\ul{#2}}}}

\newcommand{\toolLink}{\asLink{https://prong-editor.netlify.app/}{prong-editor.netlify.app}}

\newcommand{\osf}{\asLink{https://osf.io/rvyjp/}{osf.io/rvyjp}}

\definecolor{cogColor}{HTML}{ca2c92}
\newcommand\dimension[1]{\textcolor{cogColor}{\emph{#1}}}

\newcommand{\parahead}[1]
{%
  \paraheadd{#1}.
}

\newcommand{\paraheadd}[1]
{%
  \vspace{0.07in}%
  \noindent%
  \textbf{\textit{#1}}%
}

\def\subsubsec#1
{\subsubsection{#1}}

\newcommand{\inlineFig}[1]{%
  \begingroup\normalfont
  \includegraphics[height=1.2\fontcharht\font`\B]{#1}%
  \endgroup
}

\newcommand{\jsong}
{Prong}

\newcommand{\hlc}[2][yellow]{{%
      \colorlet{foo}{#1}%
      \sethlcolor{foo}\hl{#2}}%
}
\newcommand\qt[1]{\hlc[Periwinkle!15]{``#1''}}

\newcommand{\pxx}[1]{\textbf{P$_{#1}$}}

 \def\BibTeX{{\rm B\kern-.05em{\sc i\kern-.025em b}\kern-.08em
    T\kern-.1667em\lower.7ex\hbox{E}\kern-.125emX}}
\begin{document}

\title{Projectional Editors for JSON-Based DSLs
}

\author{
  \IEEEauthorblockN{Andrew McNutt}
  \IEEEauthorblockA{
    \textit{University of Chicago}\\
    mcnutt@uchicago.edu}
  \and
  \IEEEauthorblockN{Ravi Chugh}
  \IEEEauthorblockA{
    \textit{University of Chicago}\\
    rchugh@uchicago.edu}
}

\maketitle

\begin{abstract}
  Augmenting text-based programming with rich structured interactions has been explored in many ways.
  Among these, \emph{projectional editors} offer an enticing combination of structure editing and domain-specific program visualization.
  Yet such tools are typically bespoke and expensive to produce, leaving them inaccessible to many DSL and application designers.

  We describe a relatively inexpensive way to build rich projectional editors for a large class of DSLs---namely, those defined using JSON.
  Given any such JSON-based DSL, we derive a projectional editor through
  (i) a language-agnostic mapping from JSON Schemas to structure-editor GUIs and
  (ii) an API for application designers to implement custom views for the domain-specific types described in a schema.
  We implement these ideas in a prototype, \jsong{},
  which we illustrate with several examples including the Vega and Vega-Lite data visualization DSLs.
\end{abstract}


\section{Introduction}

Projectional code editors~\cite{Fowler08Projectional} combine \emph{structure editing} with \emph{rich, alternative views} of code fragments or the runtime value they produce.
Structure-editing a program's abstract syntax tree (AST) avoids various classes of errors that arise when editing text, such as typos or misremembered property names.
Custom views, sometimes called \emph{projections}, further improve usability via contextual visualizations by providing representations of code relevant to a given task, such as runtime information or execution traces.
Projectional editors then aim to make programs simpler to read and modify.

When designed for particular languages or applications, projectional editors can offer extremely rich~\cite{horowitz2023live} interactions compared to text.
For example, Sketch-n-Sketch offers both a structure editor~\cite{sns-deuce} and a direct-manipulation drawing interface for modifying and refactoring programs that generate SVG images~\cite{hempel_sketch-n-sketch_2019}.
While powerful, such editors are bespoke, requiring substantial engineering effort such that building, or even customizing, them may be out of reach to many developers.

In contrast, language-agnostic editing tools seek to augment languages in general, such as the autocomplete and refactoring tools available in editors like VSCode.
Compared to ``high-cost'' designs, such ``low-cost'' general-purpose editors are naturally less feature-rich and (by their agnosticism) unable to support domain-specific nuances.

\begin{figure}[t]
  \centering
  \includegraphics[width=\linewidth]{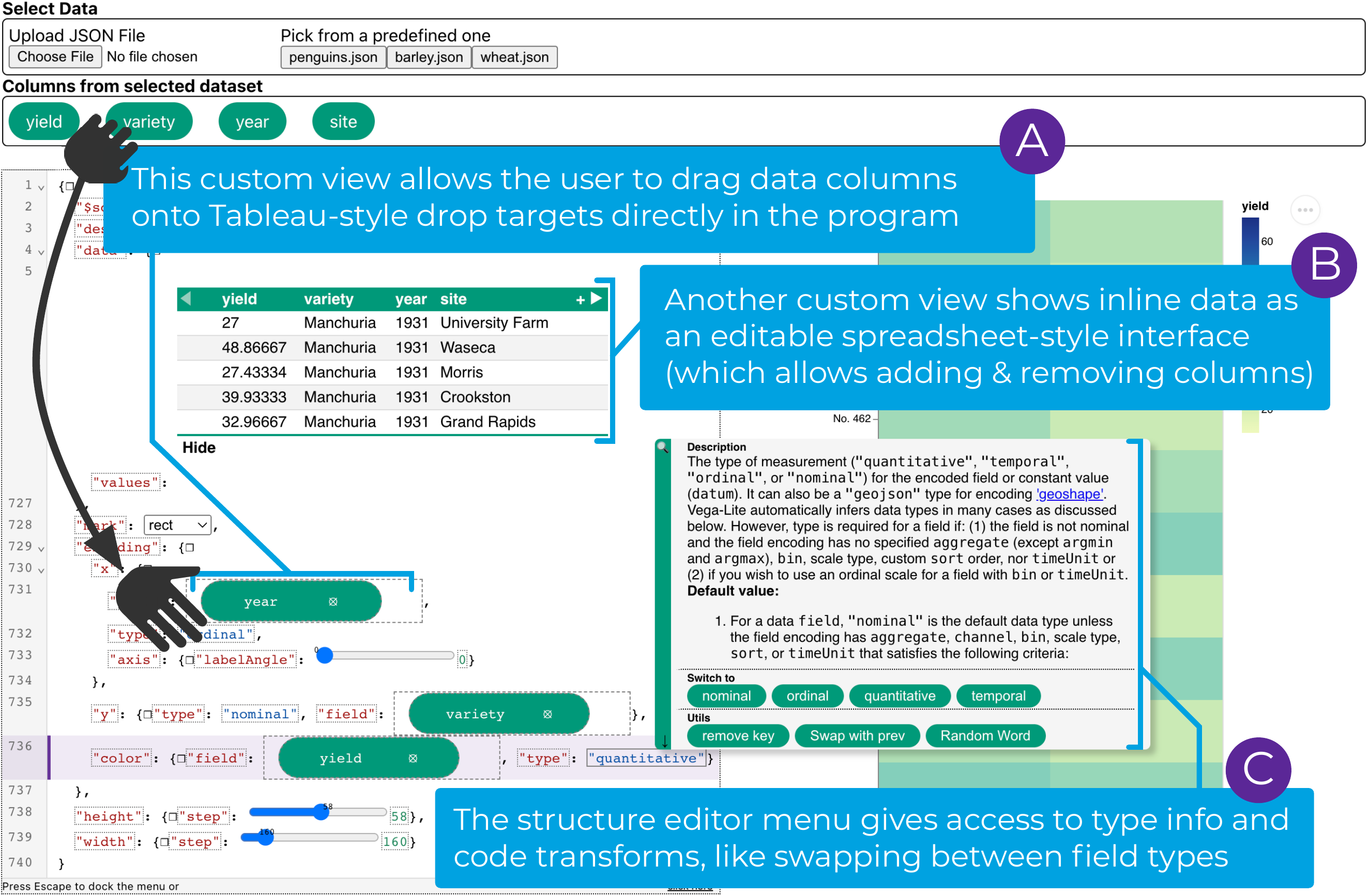}
  \caption{
    \jsong{} can be adapted to any JSON DSL with a schema. Here it is tuned to Vega-Lite via a custom Tableau-style drag-and-drop view and a view that replaces inline data with an editable spreadsheet-style display.
    Here, and throughout, figures are optimized for online zoomable viewing.
  }
  \label{fig:vega-lite-example}
  \vspace{-1em}
\end{figure}

We explore a sweet spot in this spectrum via a generic approach to designing projectional editors for a large set of diverse languages that support domain-locality without enormous engineering expense.
Namely, we consider domain-specific languages (DSLs) whose syntax takes the form of JavaScript Object Notation (JSON).
This seemingly exotic form has experienced a sudden popularity, arising in domains including
databases~\cite{mongodb_inc_mongodb_nodate},
visualization~\cite{mcnutt2022noGrammar},
twitter bots~\cite{compton2015tracery}, and
others~\cite{Varv22Borowski, payne2021danceon}.
We suggest that a suitably general and lightweight projectional JSON editor could be reused and adapted to various JSON-based DSLs for little cost---being both cheap to learn (conceptually simple) and cheap to use (short programs).

To address this goal, we present a prototype system, called \jsong{} (\underline{Pro}jectional JS\underline{ON} \underline{G}UI), that embodies two low-cost strategies (\cf{} \figref{fig:vega-lite-example}).
First, to implement a general structure editor for JSON, we develop a mapping from a JSON Schema~\cite{pezoa_foundations_2016} (a common component of a JSON-based DSL definition) to graphical structure editors.
Second, we define a simple API that allows application designers to concisely define custom views.
We provide ecological validity for this design via a need-finding study consisting of a small survey and a Technical Dimensions of Programming Systems~\cite{jakubovic2023Technical} analysis of a prominent JSON editor.
We demonstrate the practical expressiveness of \jsong{} by (re)creating examples across several DSLs.
Finally, we analyze our design's effectiveness via the Cognitive Dimensions of Notations~\cite{green1989cognitive}.
See \toolLink{} for a demo and \osf{} for code.

Through this work, we demonstrate that projections can be constructed cheaply for various DSLs.
In doing so, we point toward a future where DSLs of any type can be thriftily augmented through contextual views and visualizations.

\section{Related Work}

\newcommand{\relatedWorkHighlight}[1]
{\textbf{\emph{#1}}}

We build on prior work that improves programming interfaces via \relatedWorkHighlight{structure editors}, \relatedWorkHighlight{rich alternative views} of code, and those aiding \relatedWorkHighlight{JSON DSL usability}.

Early \relatedWorkHighlight{structure editors}, such as interlisp~\cite{interlispHistory} and the Cornell Program Synthesizer~\cite{cornellProgramSynthesizer}, were motivated in part by performance concerns: maintaining well-formed program structures throughout the development process enables the modification of running programs without having to restart execution.
Many subsequent designs (\eg{} block-based editors like Scratch~\cite{resnick2009scratch}) emphasize learnability: structure editors avoid syntax errors and facilitate syntactic and functional discoverability.
The Cornell Program Synthesizer and MPS~\cite{pech2013jetbrains} are notably  language-agnostic, allowing tool builders to specify language- and domain-specific definitions.
Modern designs\cite{pech2013jetbrains, tylr} aim to emulate familiar text editing across structural boundaries.
Others~\cite{sns-deuce, voinov2022forest} more closely resemble mainstream refactoring and autocomplete UIs, which augment unrestricted text editing with structural interactions.
More recently, Sandblocks~\cite{Beckmann23Sand} enables derivation of structure editors from arbitrary grammars while allowing some text editing actions.

Some editors render \relatedWorkHighlight{rich, alternative views} of programs and their runtime values or dynamic behavior.
Even when used solely for display (but not editing), these \emph{program visualizations} are sometimes called ``projections'' and generally fall within the goals of projectional editors~\cite{Fowler08Projectional};
examples include \emph{live programming} environments, such as Projection Boxes~\cite{lerner2020projection}, in-situ visualizations~\cite{hoffswell_augmenting_2018}, and \emph{i}-\LaTeX~\cite{gobert2022latex}.
Other editors \cite{awesomeStructureEditors, ko2005citrus, ko2006barista, omar2012active, andersen2020adding, ferdowsifard2020small} allow custom GUIs to edit programs.
For example, by dragging number sliders and color pickers, or specifying input and output examples for a desired code fragment to be synthesized.
Among such editors, Hazel~\cite{omar2021filling} notably gives custom GUIs access to the runtime produced by evaluating a partial program; this information may help the user decide how to fill in a missing code fragment.

Some tools have sought to improve the \relatedWorkHighlight{usability of JSON-based DSLs} (without resorting to hiding the underlying program as common in end-user programming tools~\cite{Zong21Lyra2,dataworld_chart_nodate, hoffswell2020techniques}).
JSON Editor~\cite{jsoneditor} and jet~\cite{jetPenner} provide structure-editing and graphical elements for manipulating JSON but do not offer additional affordances for particular DSLs.
Other JSON editors are tuned to specific use cases.
For instance, Vega Editor~\cite{noauthor_editoride_nodate} provides custom support for debugging Vega and Vega-Lite programs.
In-situ sparklines have been explored for program visualizations~\cite{hoffswell_augmenting_2018} and reactive web applications~\cite{LittRunTimeSparklines}.
Similarly, Vaguely~\cite{Travers21Vaguely} is a block-based editor used for authoring Vega-Lite programs.
Workbench-style tools~\cite{chavarriaga2023approach, petitpierre2011bottom} seek to improve the process of creating JSON DSLs.

In relation to the above characteristics, our design for \jsong{}:
is motivated by enhancing usability and discoverability;
supports only JSON syntax but is agnostic between the many DSLs embedded in JSON and specified with JSON Schemas;
enables the definition of custom GUIs, with access to live value information, for visualizing and modifying code; and
extends ordinary text editing with projectional features.

\section{Need finding study}
\label{sec:need-finding}

To understand the places where projectional-editor style modifications might usefully aid the process of using, composing, and debugging programs written in JSON-based DSLs, we conducted an analysis based on the Technical Dimensions of Programming Systems (TDPS) \cite{jakubovic2023Technical}.
TDPS is a closely-related theory to the Cognitive Dimensions of Notations~\cite{green1989cognitive}, but instead focused on the system surrounding the notation rather than the notation itself.
We use this theory to structure a close reading of a particular editor~\cite{noauthor_editoride_nodate} that exemplifies properties found in a similar system. We then use the identified issues and drawbacks to guide \jsong{}'s design.
TDPS consists of 7 clusters of dimensions (which shape our discussion), however we forgo consideration of \emph{complexity} and \emph{conceptual structure}  as they have less bearing on the design of the editor and more on the design of the notation.

To provide ecological validity to this reading, we augment this analysis with results of a small online survey with users of JSON-based DSLs ($N=10$).
Participants (denoted \pxx{X} and \qt{quoted}) were asked questions about any JSON DSL, but most discussed Vega or Vega-Lite---likely due to selection bias.
Most participants  were 30-35 years old (5/10) and male (9/10). All held degrees in computer science fields and $\geq$6 years programming experience.
They were paid via a raffle with an expected value of \$6.50.
See supplement for survey instrument (\osf{}).

\begin{figure}[t]
  \includegraphics[width=\linewidth]{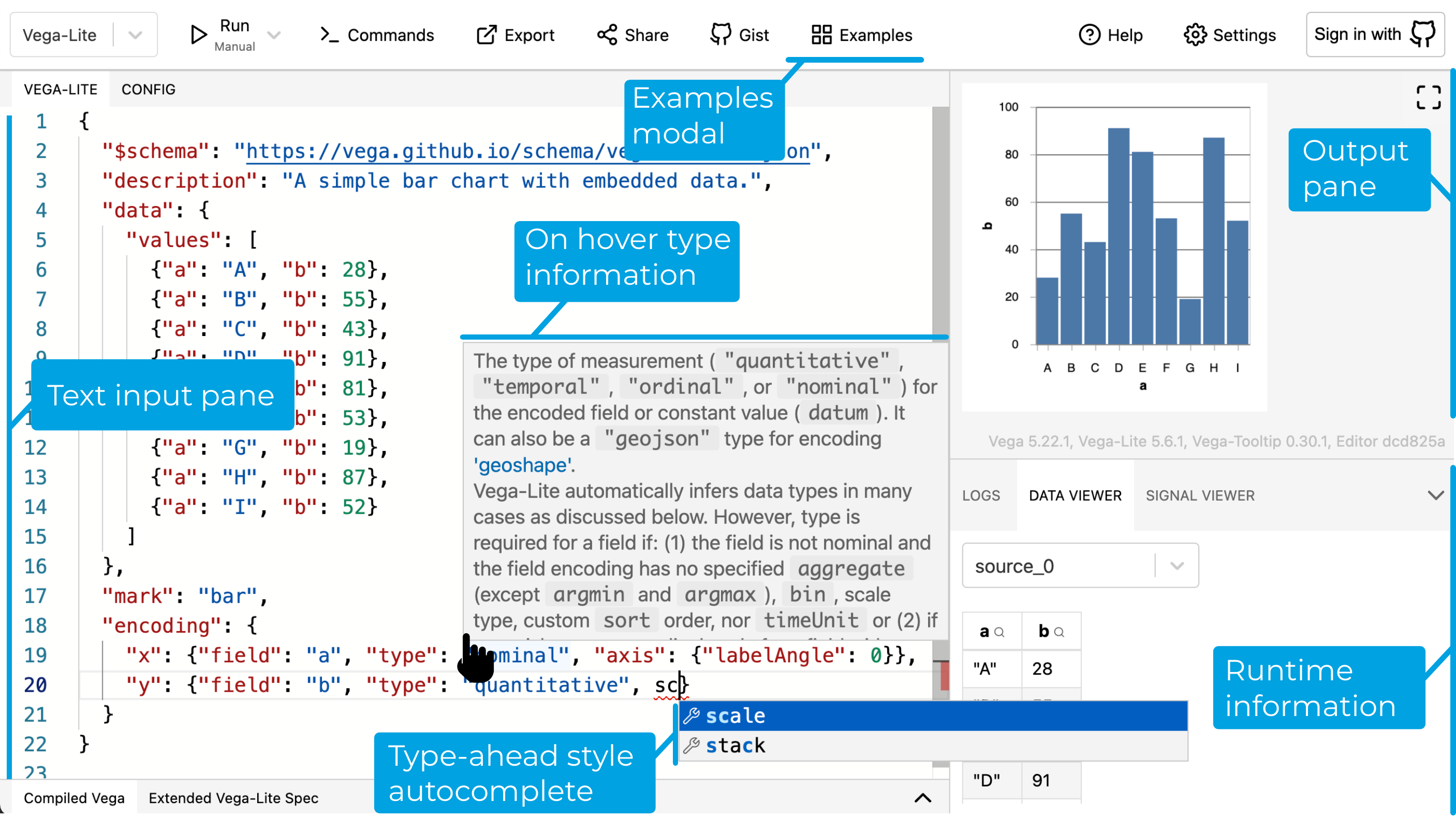}
  \caption{The Vega editor includes common functionality found in other JSON editors, so we use it to study it as a representative of this editor style.
  }
  \label{fig:annotated-vega-editor}
  \vspace{-1em}
\end{figure}

Our analysis considers the Vega Editor~\cite{noauthor_editoride_nodate} (\figref{fig:annotated-vega-editor}).
We focus on it because it includes most features found in similar editors and thus can be seen as representative of the form.
It features various language server protocols (LSP)-based capabilities~\cite{vscodeJSONLSP}, such as simple autocompletes, introspection, and so on.
These features are found in similar editors (but adapted to their domains), including deneb's BI-centric charting~\cite{deneb}, MongoDB Compass's~\cite{mongod_compass} NoSQL aggregation query building, and even Tracery Editor's~\cite{traceryEditor} support for generative text programs.

\parahead{Interaction}
\label{sec:interaction}
The first TDPS cluster considers the way in which users manifest their ideas, such as through different sizes of feedback loops or gulfs of execution.

The primary interaction loop in the Vega Editor involves textual input, viewing the created visualization, and then modifying it through further textual input.
Within this loop are conventional text editor interactions, including linting (providing semantic validation),  autocomplete (providing a simple syntactic search), and hover type hints (providing code explanation).
Outside of this loop are larger cycles that involve debugging (such as by consulting the data, signal viewer, or developer console), consulting the documentation, and specifying the data.
The last of these might require tedious round trips to make minor updates to the data.
Vega and Vega-Lite approach this issue via a data transformation notation, however debugging this can be opaque and difficult for the user. For instance, \pxx{7} notes that it is \qt{easier for me to do data processing outside of Vega.}
We address these issues by supporting \emph{in-situ} projections of the state within the text, as in the inline data table in our Vega-Lite case study (\figref{fig:vega-lite-example}).

The presence of a common loop at different granularities suggests that interventions could be made to tighten the loop bounds.
For instance, moving documentation look up into one of the interior loops, such as by structure editing, has clear value. Moreover, it seems to be a feature that users already desire.
For instance, \pxx{3} wanted a way to know all \qt{the available values for each thing}, and \pxx{5} noted that \qt{I can never remember the syntax of anything. Where does ``mark'' go, where does ``dataset'' go or is it actually ``data'', is it ``dot'' or ``point'' mark etc.}
Autocomplete can address some of these issues, but its design limitations can lead
it to \qt{not always work} (\pxx{6}) and be unable to help place arguments. We explore an alternative approach in our \emph{in-situ search} case study.
Prior work~\cite{sns-deuce} has shown the value of mixed textual and graphical structure editing as it allows the user to make modifications as they wish, without being forced into a single mode.
This notion guides the polite~\cite{whitworth_polite_2005} design of our structure editor.

\parahead{Notation}
\label{sec:notation}
The next TDPS cluster considers the notations available to the user of the system.
Three top-level notations exist within the editor, Vega, Vega-Lite, and a CSS-style styling language (for setting colors, fonts, and so on).
These notations are housed within JSON, which has usability issues of its own. For instance, the lack of comments (\pxx{2}, \pxx{3}) and the syntax can be fiddly (such as due to \qt{trailing commas} \pxx{3}).
Reducing this \dimension{error-proneness}~\cite{green1989cognitive}, such as by simplifying  symbol input,  motivates our use of a structure editor.

Various minor notations are scattered throughout these DSLs. These include the Vega Expression language (which is used for controlling Vega's variables), web-driven properties (such as the hex color language), as well as more domain-specific elements such as the d3 axis format language.
Some of these are more full-featured than others.
For instance, \pxx{7} noted that the \qt{expression language is basically a DSL within a DSL.}
\pxx{2} desired \qt{syntax highlighting for the expression language.}
Similarly, \pxx{4} observed \qt{I think one limitation is that sometimes properties can hold strings and objects.}
We address these issues by creating means for alternative interfaces for these languages through views.

\begin{figure*}[tb]
  \centering
  \includegraphics[width=\linewidth]{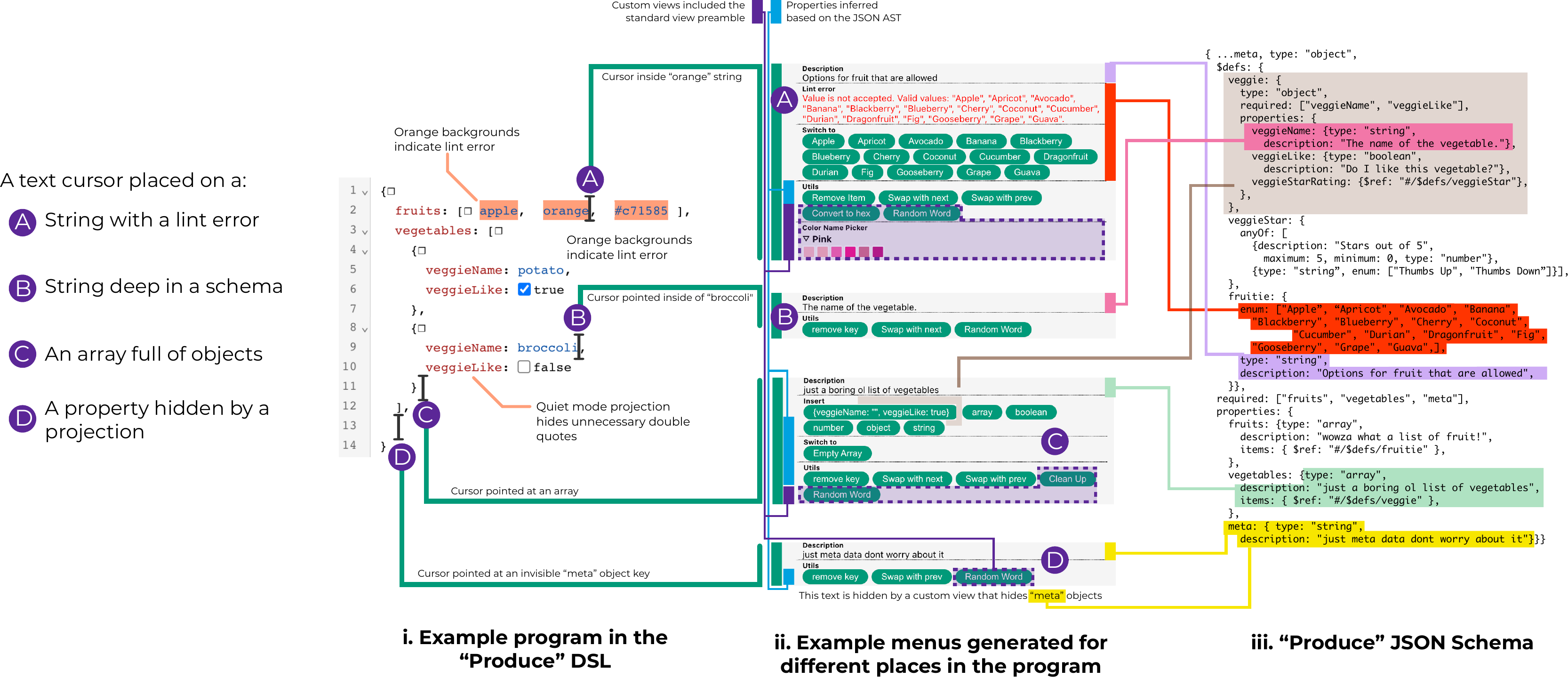}
  \caption{An illustration of the mapping between the program, structure editor menu, and JSON Schema for a toy ``produce'' DSL.
    (i) Some cursor positions that generate some menus (ii). (iii) how each menu element was generated---\ie{} via the schema or various views.
  }
  \label{fig:big-schema}
  \vspace{-1em}
\end{figure*}

\parahead{Customizability}
\label{sec:customizability}
This cluster considers how the user can modify the usage of the system to fit their interests and how system state can be manipulated.
Generally, the editor can not be modified by the language or even much beyond some limited aesthetic options.
Similarly, state is viewable through the signal and data viewer, but can not be changed. \pxx{7} noted that \qt{it is hard to get info out of the runtime} but that the \qt{data viewer and signal viewer are very useful because they give visibility to state of runtime.}
Surfacing tools that allow the user to modify the user experience may allow them to specify things in useful ways.
To this end, bidirectional editing of state and code (such that modifying state updates the code that generates it) seems especially useful.
This is explored in Chart Studio~\cite{chart_studio_plotly} and our Tracery case study.

\parahead{Errors}
\label{sec:errors}
The penultimate cluster of concern considers how errors are surfaced through the system.
The editor can detect syntax errors (such as incorrect JSON syntax), some semantic errors (such as using properties that are not allowed by the JSON Schema), and runtime errors (as arising from Vega programs execution).
Dealing with these error types can be difficult~\cite{hoffswell_visual_2016}: \qt{Debugging can be hard.  When things go wrong, you often just get a blank output} (\pxx{2}).

The Vega Editor acts most concretely on the first two types.
Syntax errors block execution and are highlighted in the input pane.
Runtime errors may stop execution depending on the severity and are shown in the system information area.
The support might be expanded by, as \pxx{5} suggests, \qt{looking at browser dev tools could be an option,} adding that \qt{Firefox, e.g., tells you which of your CSS rules don't have an effect....that would be cool for vega, too.}
We design our view system so that prior bespoke works (such as those on debugging Vega~\cite{hoffswell_augmenting_2018}) are simpler to form and modify.

In contrast, the editor does little to help the user deal with semantic errors, primarily due to limitations in the Vega and Vega-Lite JSON schemas.
For instance, \pxx{1} observed that based on current support \qt{it's not clear which properties are valid for different chart types.}
Such issues could be addressed by redesigning the JSON Schema to better capture such issues, or additional layers of tooling could be introduced (as in visualization linters~\cite{mcnutt_surfacing_2020,hopkins_visualint_2020, chen2021vizlinter,mcnutt_linting_2018}).
Our system is designed to provide simple support for adding additional validators.
As our malleable views power this support,
it is not bound to highlighting errors via
\colorwave[red]{wavy underlines}.
See \figref{fig:big-fig}A or \figref{fig:tracery}.

\parahead{Adoptability}
\label{sec:adoptability}
The final TDPS cluster studies how the system helps or hinders adoption by various users.

The learning curve of JSON-based DSLS can be high for new users~\cite{mcnutt2022noGrammar}.
This is likely due to the often off-putting JSON-based syntax, and, as noted by \pxx{6}, \qt{the toolchain for non-js coders is quite burdensome.}
The editor does help with the latter of these issues by making it easy to quickly drop into a chart, however, it can still be challenging for novices, which has led to a variety of tools that seek to make them easier to use~\cite{Travers21Vaguely,mcnutt2021integrated, guzdial2021integrating}.
Supporting novices by providing system-level support for simplifying the idiosyncrasies of JSON programming informs the design of our structure editor.

A key way in which novices are supported in the Vega ecosystem is through a close binding between the documentation and the editor in a way that supports opportunistic programming~\cite{brandt_two_2009}.
For instance, the editor's examples modal exposes the hundreds of chart types (mirrored from the documentation), acting as something akin to a chart chooser~\cite{grammel_survey_2013}.
This allows users to jump to particular charts and then adjust them to task---although, as explored in Ivy~\cite{mcnutt2021integrated}, tooling can further simplify this process.
To this end, \pxx{4} noted, \qt{so I take an example that is closest to what I want to do and then proceed from there.}
Within the notations, a wide array of functionality is provided, capturing numerous common data transformation operations (among others).
Yet, \pxx{4} noted that the data transformation language is \qt{hard to grasp too and the data viewer could do a better job in communicating these.}
\pxx{10} observed \qt{there's a lot of hidden state then requires guessing.}
The editor has some support for this type of inquiry, but it is inadequate for some users---\pxx{8} wished that the data view \qt{was more insightful, intuitive, and easier to explore.}
We seek to close the gap between context and text by designing our views so that JSON DSL novice-friendly UIs are simple to build.

\section{System: \jsong{}}
\label{sec:system}

To explore how we might address the limitations of prior JSON editors, we developed \jsong{} (\underline{Pro}jectional JS\underline{ON} \underline{G}UI).
This modular framework supports JSON DSL use by enabling browser-based projectional editing in the context of a full featured text editor (CodeMirror~\cite{codemirror}), consisting of two components: an automatically inferred structure editor (based on a JSON Schema~\cite{pezoa_foundations_2016}) and custom views that commingle with text (provided by the application designer).
These tools enable cheap domain-specific editors in a practical setting
that previously might have required protracted engineering effort.

\begin{figure*}
  \includegraphics[width=\linewidth]{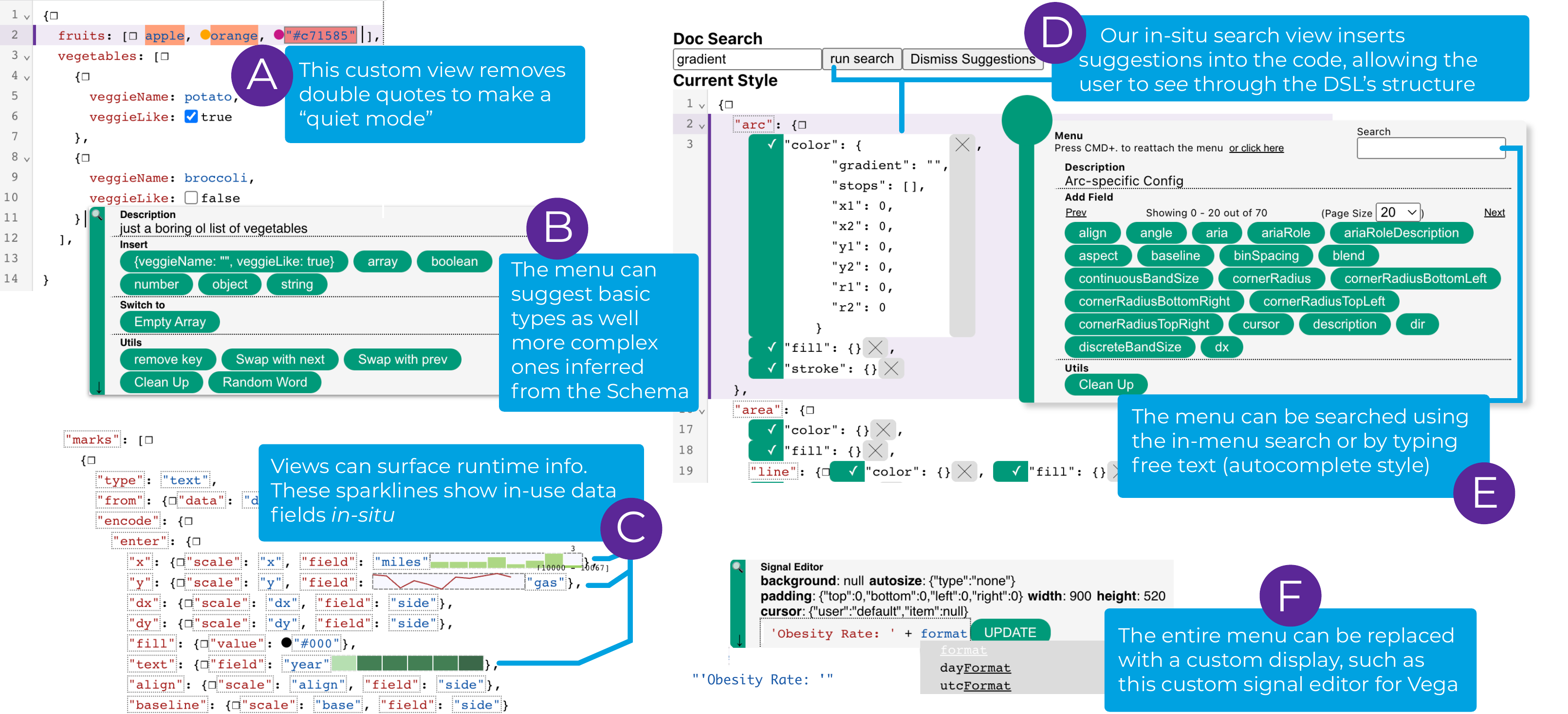}
  \caption{
    \jsong{}'s projectional editing consists of schema-derived structure editing (B, E) and custom views (A, C, D, F).
  }
  \label{fig:big-fig}
\end{figure*}

\parahead{Structure Editing}
\jsong{} enhances text manipulation through a dynamic menu that provides type information, structural manipulation, and DSL-aware suggestions.
The basic interface is that a of text editor, so the underlying source of truth is text, but additional means of editing (namely via our menu).
It follows the text caret as in an IDE-style tooltip (\figref{fig:big-fig}A), appears docked at the bottom of the editor, or floats in a user-specified location (\figref{fig:big-fig}E)---allowing the user to decide how much visual response fits their usage.

The simplest controls in the menu come from JSON parse-tree data types. This enables basic text manipulations (\eg{} inserts, deletes, and reorders) to be conducted graphically. This limits some of JSON's well-known \dimension{error proneness}.
More critically, however, is the information derived from JSON Schemas~\cite{pezoa_foundations_2016}.
Schema are JSON objects that describe allowed forms for a given JSON DSL---analogous to providing TypeScript types for an untyped JavaScript (JS) library.
While JSON DSL usage is often supported by schemas~\cite{noauthor_editoride_nodate}, we demonstrate how those schemas can be used to form structured editing elements---\eg{} buttons that insert synthesized objects (by identifying required props) or option picklists---instead of merely for validation~\cite{ajv} or autocomplete~\cite{schemaStore}.
We compute applicable types for every AST node (via the schema) and then apply them to a suite of heuristics to generate all relevant information and potential changes from the current state (\cf{} \figref{fig:big-schema}).
For instance, the user in \figref{fig:vega-lite-example}C can click buttons for each allowed field type
(\emph{nominal}, \emph{ordinal}, \etc{})
rather than needing to remember or type them.

To simplify searching the structure editing options, \jsong{} includes a simple autocomplete that uses incomplete strings as a signal to filter the menu.
This coarse autocomplete captures many of the same properties as more complex recommenders without requiring the weighty configuration of tools like a language server.
Autocomplete~\cite{schemaStore} is typically available for JSON DSLs via Schema support~\cite{schemaStore}, however, \jsong{} is unusual in its support of browsing all possible values for a given position, which supports our respondents' desires to that effect.
Prior work~\cite{oney2012codelets, rong2016codemend} has explored some related designs.

\begin{figure}[t]
  \small{}
  \newcommand{\myBump}{\hspace{0.08in}}
  \newcommand{\lineSep}{{\hspace{0.04in} | \myBump }}
  $$
    \begin{array}{ll}
      View  & =\Bigg\{
      \begin{array}{l}
        {placement} = `inline'\myBump|\myBump`replace'\myBump|\myBump`menu' \\
        {query} = Query,                                                    \\
        {content} = React Component                                         \\
      \end{array}                         \\                    \\
      Query & = (SyntaxNode, ASTNodeName[])             \\
            & \lineSep{} (KeyPath, keyPath[])           \\
            & \lineSep{} (SchemaNode, SchemaNodeName[]) \\
            & \lineSep{} ...
    \end{array}
  $$
  \normalsize{}
  \vspace{-1em}
  \caption{Formal description of our views.
    SchemaNode refers to the inferred type of a given node from the Schema, whereas SyntaxNode refers to the type inferred from parsing.
    See \toolLink{} for implementation-specific details, such as component props.
  }
  \label{fig:view-eqns}
\end{figure}

\parahead{Custom Views}
\jsong{} supports domain-specific functionality through views.
These modular specifications let the application designer  describe how and where custom components should be placed, enabling construction of numerous potentially bespoke UIs in a systematic manner.
For instance, sparklines\cite{goffin2016exploratory} can be used to show runtime data inline~\cite{hoffswell_augmenting_2018}, or a spreadsheet might replace a block of tabular data---examples we explore in our Vega case study.
While not prescribed, the intent of views is to present DSL information in a fashion that might more effectively match that content.
This design seeks to address the limitations found in conventional editors due to their \emph{unrich} connection with their output, which often leads to difficulties in both text authoring and debugging.

As formalized in  \figref{fig:view-eqns}, views are {placed} \emph{inline} with the text (as prefixes, suffixes, or backgrounds), \emph{replace} the text, or appear in the \emph{menu}.
{Queries} executed during a syntax tree traversal determine on which AST nodes views will appear.
Queries are a tuple consisting of the query content and its type, which might include the location in the syntax tree (\emph{KeyPath}), the inferred data type from the JSON Schema (\emph{SchemaNode}), the AST Node type (\emph{SyntaxNode}).
For instance, the query \emph{(SchemaNode, exprString[])} selects all Vega Expressions in a Vega-Lite program.
Views follow a CSS-style cascade, so that later views take priority over earlier ones. For instance, two \emph{replace} views whose queries match the same node,
the one specified later will be used.
View {content} consists of a React component which can contain any details present in the containing application, or automatically inferred data, such as types from the schema.
For instance, runtime info can be integrated directly into the text via sparklines.
Views' unprescribed format supports numerous designs while offering a familiar API for many developers (conceptual thrift).

Views are supplemented by a preamble of common tools (as in JSON Editor~\cite{jsoneditor}), including boolean toggles\inlineFig{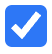}, color indicators ${\color{red}\bullet}$, color pickers, number sliders\inlineFig{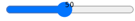}, code formatters, and object sorters.
These elements can be removed by the application designer if they are unnecessary.

\section{Case Studies}
\label{sec:eval}

We next demonstrate the expressiveness of this approach via several case studies.
We focus on these DSLs because they are intended for human use~\cite{mcnutt2022noGrammar}, have differing semantics, and have well-formed schemas.
Demos are at \toolLink{}.
\footnote{These demo implementations make some functional simplifications to show the properties of concern (\eg{} the Vega-Lite example only supports small data). However, because of \jsong{}'s architecture, each DSL is fully supported---although further projections are necessary to use them as a complete end-user application.}

\parahead{Warm-up}
\label{sec:warm-up}
To begin, consider a toy example.
JSON can be verbose, as it features numerous double quotes for string literal and object properties.
To reduce this visual noise, we can construct a quiet view that prunes double quotes:

\small{}
$$
  \left\{
  \begin{array}{l}
    target = (SyntaxNode, [PropertyName, String]), \\
    context = `inline'                             \\
    content =                                      \\\hspace{0.2in}\verb+(props) => <div>{trim(props.value)</div>+               \\
  \end{array}         \right.                            \\                    \\
$$
\normalsize{}
This small amount of code transforms the editor (\figref{fig:big-fig}A) without requiring heavyweight extension code used in tools like base CodeMirror~\cite{codemirror} or VSCode.
For instance, a minimal React application enabling quiet mode requires 128 lines of code (LOC) in base CodeMirror and only 42 LOC in \jsong{} (\asLink{https://prong-editor.netlify.app/\#/quiet-mode}{see link}).
Verbose extension APIs (like CodeMirror) support the construction of any extension for any language and text, whereas ours augments the JSON AST.
This exchange is one of domain-specificity: by simplifying the API, we make many editors simpler to produce while making some impossible.

\newcommand{\tool}[1]{\textsc{#1}}

\parahead{In-situ Visualization with Vega}
\label{sec:in-situ}
Vega is famously difficult to debug~\cite{hoffswell_augmenting_2018, hoffswell_visual_2016} as it possesses ample \dimension{hidden dependencies}, the navigation of which can yield \dimension{hard mental operations}.
\footnote{
  These issues are well enunciated by the Cognitive Dimensions of Notations~\cite{green1989cognitive},
  which is a set of heuristics used to structure close readings of interfaces.
  We reference these \dimension{dimensions} throughout to note usability issues.
}

An exciting approach~\cite{hoffswell_augmenting_2018, LittRunTimeSparklines} to these issues has been to add visual presentations of these \dimension{hidden dependencies} in the code itself---such as through sparklines, like \inlineFig{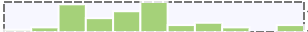}.
Yet such enhancements are typically bespoke, requiring expertise with the extension environment, putting them far beyond many developers.
\figref{fig:big-fig}C shows a view that inserts several families of sparklines showing univariate data summaries.
Our sparklines extend prior work~\cite{hoffswell_augmenting_2018,LittRunTimeSparklines} by allowing users to toggle through different visualizations, rather than fixing a single one.
This is enabled by our lightweight component architecture.

Debugging Vega is made more difficult by the DSLs embedded  in its schema.
For instance, it includes an expression language (a JS subset specified in string literals) that controls its variables (called signals).
This supports a range of expression types that would be difficult to capture via JSON-based operators,
for instance, the snippet {\small{}\texttt{"1 + sqrt(datum.size) / 2"}} concisely describes a simulation force.
Yet, editors typically do not provide syntax highlighting for string-embedded DSL or other typical editor affordances like autocomplete, which can yield an \dimension{inconsistent} interaction between the notations.
These issues impede reasoning about whether a given expression is correct.
To address these issues, we created a signal editor view (\figref{fig:big-fig}F),
which includes a syntax highlighting code editor with a Vega Expression-aware autocomplete.
To compose a new signal, the user navigates to an expression and starts typing in the menu, receiving autocompletes and validations (\ala{} linting) of their expression.
By supporting nested DSL usage, we can reduce some of the cognitive burdens of using Vega.

With relatively little work, we can (re)create sophisticated bespoke editors and resolve issues latent to this form of DSL.

\parahead{Blurring Text and GUI in Vega-Lite}
\label{sec:vega-lite-case-study}
Vega-Lite simplifies Vega by limiting what is expressible in favor of a syntax tightly bound to the grammar of graphics.
Yet, its sometimes \dimension{terse} and \dimension{secondary notation}-free syntax can be difficult for novices~\cite{mcnutt2021integrated, naimipour2020engaging, nelson2021creating}.
Previous solutions often put a facade between the user and text, letting them interact indirectly with the DSL via a GUI---as in Lyra~\cite{Zong21Lyra2} or Voyager~\cite{wongsuphasawat_voyager_2017}.
Here we explore how \jsong{} can help weave GUI interactions into text editing.

\begin{figure}[t]
  \includegraphics[width=\linewidth]{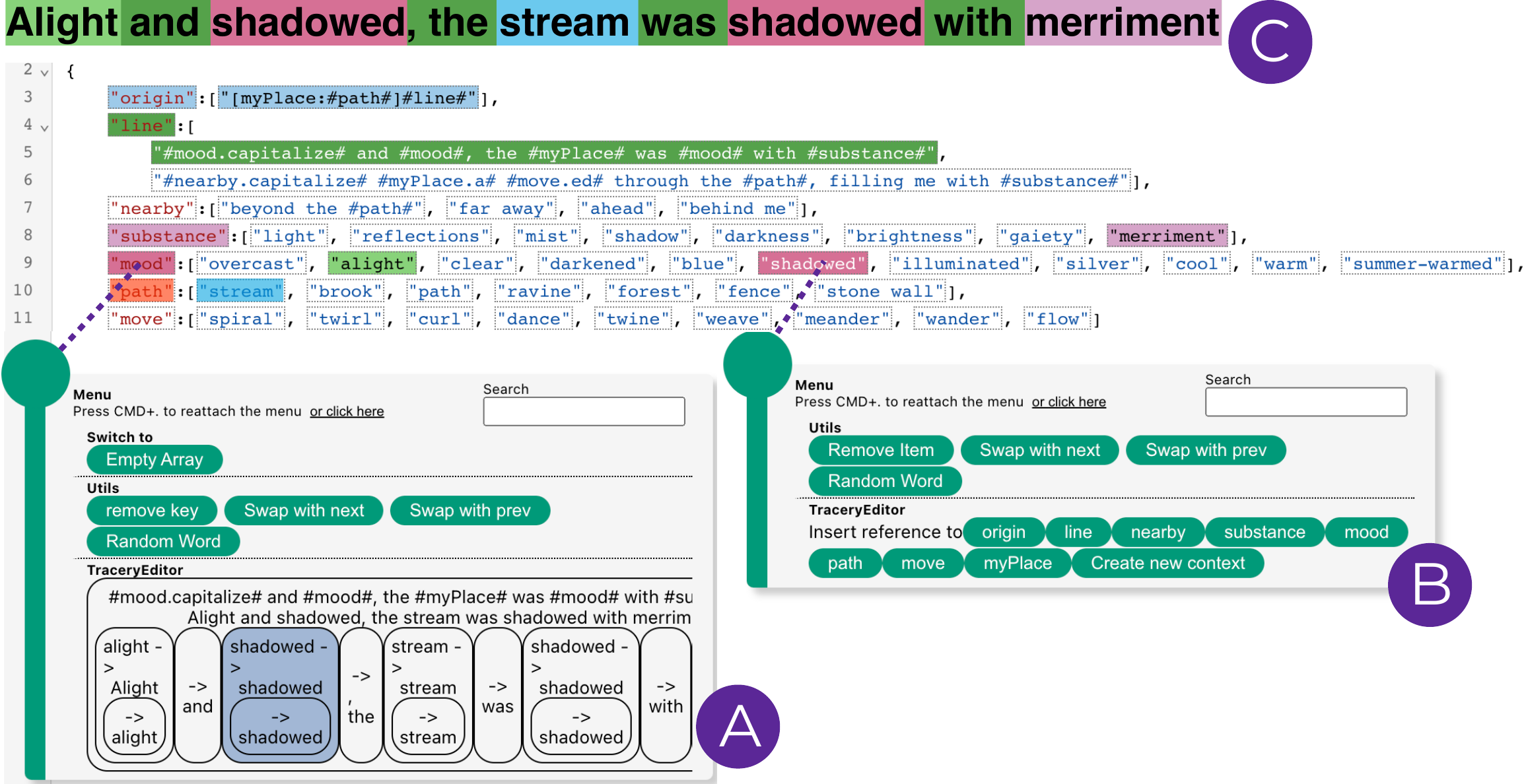}
  \caption{\jsong{} supporting the generative text language Tracery~\cite{compton2015tracery}.
    This instance has simple bidirectional editing between the output and input, and syntax highlighting for the specific values involved in the output.}
  \label{fig:tracery}
  \vspace{-1em}
\end{figure}

We explore the potential of this approach through several views.
First: a Tableau-style drag-drop interaction, in which data columns are dragged from an automatically inferred set of ``pills'' onto ``shelves'' that replace text that describes  the binding of data fields to graphical channels.
For instance, to change the encoding of the x dimension in \figref{fig:vega-lite-example}A the user drags one of the data fields above the editor onto the relevant field target.
When the text caret enters the drop targets, they switch from GUI to text---allowing more familiar users to use text if they wish (functionality that can be activated per view).
Next, we replace inlined data (arrays of objects) with a more graphically approachable editable data table (\figref{fig:vega-lite-example}B), which operates akin to a familiar spreadsheet interface.
Views can be introduced to modify the program to better align with the form the application designer wishes to use.
For instance, we include a view that converts data references into inline data (which is better suited to our data table component's expectations) and another that allows the user to upload data into a specification without copying and pasting.
Finally, we add pick lists directly into the UI so that users might more easily select values relevant to them, such as allowing the user to select a mark type.
Such views can simplify JSON-based DSLs use, as they provide a straightforward way to use and modify a potentially unfamiliar DSL.
Vega-Lite features a well-annotated JSON Schema that enables our structure editor to provide useful suggestions and options.
Yet this schema is imperfect and can make it difficult to check if a chart is being used in a semantically incorrect way.
For instance, the \emph{circle} mark type does not support \emph{text} encoding, as that encoding field is unique to the \emph{text} mark, yet nothing in the schema would prevent the user from authoring such a program.
Thus, while schemas cheaply give basic validation, they do so potentially at the cost of being unable to handle some usability issues.
A custom linter~\cite{chen2021vizlinter,mcnutt_linting_2018,mcnutt_surfacing_2020,hopkins_visualint_2020} could potentially guard against issues like this, however, such tools require careful construction.
Our view system supports simple construction and presentation of such validators (such as through the CSS class-based styling used in \figref{fig:tracery}), but this places a non-trivial burden on the application designer to re-engineer the intended semantics of the DSL.

Through these modest views, we are able to create novel domain-specific interfaces that simplify interactions with a sometimes thorny or complex visualization DSL.

\pagebreak

\parahead{In-situ Search for Vega Config}
\label{sec:vega-config}
Vega Config is a CSS-like DSL for making reusable chart styles in Vega and Vega-Lite .
Creating such a style is a non-trivial task because Vega Config is closely bound to the rendering context and does not have Vega/Vega-Lite's elegant formalisms.
These factors cause it to be a large and \dimension{inconsistent} DSL.
Such \dimension{inconsistencies} can lead to frequent docs searches---cycles that can cause the user to lose context~\cite{mcnutt2021integrated} and lead to the \emph{language cacophony}~\cite{fowler2010domain} DSL failure mode.

Among such external loops, there are several types of search, including what the specific syntax of a given property is (as in Barke \etals{}~\cite{barke2022grounded} acceleration), browsing available functionality (as in their exploration), and where a property should be placed (which is more local to JSON-based DSLs).
\jsong{} supports the first two issues via the menu and autocomplete.
For placement, we create an in-situ search view that generates inline suggestions, as shown in \figref{fig:big-fig}D.
Suggestions are generated by a schema traversal that matches properties like name or description with a query string.
These are then formed into JSON snippets from the traversal path, inserted into the code, and visually marked as suggestions.
This addresses respondents concerns about knowing where in a specification to place an argument in a JSON program.

A user composing a chart style has then several entry points. They might begin by typing part of a property name to receive IDE-style autocomplete menu filtering.
They can also start by searching for a particular endpoint they wish to find, such as by typing a color scheme they want to include (\eg{} \emph{cividis}) into the in-situ search, which then will generate suggestions including each of the intermediate steps to reach their goal.
They can then click through those they want to accept and dismiss the rest.
A \emph{richer}~\cite{horowitz2023live} UI would allow user to select and style parts of a rendered chart directly, without needing to pass through text.
Yet, such an approach would require careful instrumentation of the Vega renderer.
This search could be replaced with an AI-based code-generator model~\cite{blinn2022integrative},
which might enrich the results at a steeper cost.

Our approach gives ample value for little implementation cost ($<$500 LOC) in a way that can be  applied to any schema.

\parahead{Tracery: Bidirectional Views}
\label{sec:tracery}
A key advantage of using JSON for a hosting language of DSLs is that it is simple for mechanizations to operate.
Here we explore how this property can simplify the creation of applications that are tightly synchronized between representations in Tracery~\cite{compton2015tracery}.

Tracery is a generative textual language intended to support casual creators by creating text from a predefined grammar, facilitating tools like Twitter bots or procedural narratives.
It consists of an input grammar (a JSON object describing the grammar) which is used to generate a piece of text.
The language has a type \verb+Record<Symbol, Rule[]>+ where \verb+Rule+s are strings that may contain references to symbols (\eg{} \verb+#mood#+). An instantiation is created by recursively expanding from a start symbol.
The DSL's simplicity matches its intended users, who lack technical skill (such as children~\cite{comptonTraceryMiddle}) or whose expertise lies in other domains (such as artists).

Despite the apparent simplicity of the language, its users may have trouble tracing the relationship between the program output and the grammar, yielding \dimension{hard mental operations}.
To address this, we developed an instance of \jsong{} tuned to support Tracery (\figref{fig:tracery}).
This system includes several custom views.
\figref{fig:tracery}A highlights a menu focused on one of the \verb+Symbol+s, which shows several type-based suggestions (such as remove key) as well as a Tracery Editor-inspired view~\cite{traceryEditor} (TraceryEditor) that shows the relationship between the constructed piece of text and the grammar.
\figref{fig:tracery}B shows a menu focused on a \verb+Rule+ that includes suggestions to insert references to the \verb+Symbol+s.
Throughout the text pane \verb+Symbol+s and \verb+Rule+s  are colored to highlight their usage. These colors are synchronized with the output (\figref{fig:tracery}C) to simplify tracing the connection between output and grammar.

Finally, the output and the grammar are bidirectionally synchronized, such that edits to the output are propagated back into the grammar.
In particular, edits to the output are returned to the grammar through a simplistic enumerative synthesis algorithm, in which the modification to the output is identified (delete, insert, swap). Then an appropriate modification is made to every \verb+Rule+, until the expected output is formed. If this fails (such as because a \verb+Rule+ is in use twice), then a provenance-based algorithm (based on annotating the execution tree) is used to identify the string position.
When unsure how to proceed, the system does not propagate the edit back into the grammar, instead alerting the user that it is out of sync.
While imperfect, it offers functionality not previously available for this DSL.

These views provide a rich connection between the input language and the output.
This example further highlights the simplicity of using \jsong{} as part of a UI.
While this was the most involved of the studies, requiring the most bespoke crafting, it was achievable in $<$800 LOC.

\section{Analysis}
\label{sec:cdn}

Here we analyze \jsong{} using the Cognitive Dimensions of Notations~\cite{green1989cognitive} (CDN).
We use CDN (rather than TDPS) because it can be used evaluatively~\cite{beaudouin2004designing}, whereas TDPS is primarily a descriptive theory.

\jsong{} utilizes a text-forward approach compared to other structure editors, which often preference GUI representations at the cost of losing some textual expressiveness.
In giving primacy to text, we also inherit all of the issues latent to it.
For instance, JSON's strict textual syntax can be highly \dimension{error-prone}, as even small typos will cause programs to be unreadable by many parsers, yielding a disproportionate error response.
\jsong{} addresses this issue by utilizing the forgiving JSONC parser~\cite{jsoncParser} (which supports comments as a \dimension{secondary notation}), and by offering structural edits (such as inserting new elements in arrays) so that such issues are harder to encounter.
Yet it is not impossible to encounter such issues. \jsong{} has limited error handling capabilities, whose default is to deactivate views in the presence of an unresolvable error. This may confuse the end-user, who is not clearly warned.
While primarily an implementation issue, the design for such alerting needs to be clarified.

Just as in other declarative languages, it can be challenging to debug~\cite{gathani2020debugging}, requiring the programmer to hold a potentially unfamiliar execution model held by the DSL in their mind to address possible subtle errors (providing \dimension{hidden dependency} and \dimension{hard mental operations} difficulties).
We follow Hoffswell \etals{} \cite{hoffswell_augmenting_2018} \emph{in-situ visualization} in that our custom views offer a space for the application designer to surface relevant information into the editor to help end-users debug.
However, this support does not come for free, and so it pushes the burden on the application designer to make the tool useful.
Similarly, the other common shortcomings of JSON editors  (such as difficulties with \dimension{juxtaposability})  are not addressed by \jsong{} directly.
Instead, the application designer is simply given more places to help resolve those issues.
\jsong{} seeks to improve the \dimension{closeness of mapping} between using a language and its output through its projections, but this approach can be somewhat artificial.
Enhancing the textual form of JSON-based DSLs supports their use---a task distinct from the actual task that the DSL aims to achieve:
making it easier to author Vega programs is different from presenting data.
Such \emph{rich}~\cite{horowitz2023live} interactions are ideal but difficult to achieve in a cheap generalizable manner, as in our Tracery case study.

From an application developer's perspective, \jsong{} is also not without cognitive roadblocks.
Our query-and-effect model for specifying view placement utilizes likely familiar CSS-style conventions. However, this exposes CSS-style issues, such as \dimension{progressive evaluation} difficulties in the potentially opaque placement of nodes.
Further, our AST and JSON Schema-centric mental model might be challenging to some developers.
We try to address these issues via a debugging view that exposes properties like inferred types and node key paths, but this will do little to reduce \dimension{hard mental operations} for some users.
Yet, some designs are simply impossible with our approach. For instance, our views cannot modify each other (\ala{} CSS cascades), making some context-sensitive queries difficult to produce.
Other designs (like CodeMirror's functional-update loop) could address some of these issues but induce other burdens of their own (such as a more \dimension{vicious} and complex usage) that may be unfamiliar to many.

\section{Discussion}
\label{sec:conclusion}

We have shown a way to make projectional editors for JSON-based DSLs cheaply (in terms of implementation size and conceptual burden).
This enables straightforward construction of designs that might otherwise require substantial effort.

We used our prototype instantiation of this idea, \jsong{}, to explore how text-based interfaces can be the foundation of a UI rather than merely a part of one---in contrast to tools like Ivy~\cite{mcnutt2021integrated} or GALVIS~\cite{shen2022galvis}, which attempt to enhance JSON-based DSL usage by surfacing them as a \emph{component} of the interface.
From these explorations, we suggest that contexts where DSLs (like SQL) are prominent might surface those DSLs in the UI rather than (digitally) papering over them.
For instance, a dashboard could show data provenance by weaving its charts into its SQL queries.
Numerous new application forms open by viewing the text plane as an approachable part of UI design.
However, like any work, this one has its share of limitations and areas which could be usefully expanded.

This paper focused on showing that bespoke editor enhancements can be built cheaply and systematically, rather than verifying such interventions are valuable to users.
Prior studies---both from works debuting some of these techniques~\cite{lerner2020projection, hoffswell_augmenting_2018} and those considering projectional editors more generally~\cite{berger2016efficiency, chodarev2022experimental}---provide empirical evidence of the utility of these interventions.
As a way to check these properties we employed a theory-forward evaluation that allowed us to identify various short comings.
While a useful form of critical reflection~\cite{satyanarayan_critical_2020}, we intend expanded on these considerations by exploring how useful projectional editors are for realistic lay end-users in practice, once we have refined \jsong{} for wider release.

Sul{\'\i}r \etal{}~\cite{sulir2018visual} describe a design space of visual augmentation of source code editors. While \jsong{} can produce most designs, it can only place elements on the AST, which precludes marginalia and line-based views. This is a limitation of our query language that we intend to address.
JSON Schemas provide substantial value for us but also have non-trivial limitations.
They can be challenging to author or maintain and have known design flaws~\cite{JSONSchemaIssue, baazizi2021empirical} (like being unable to validate within strings).
Growing tool support (\eg{} auto-generating schemas~\cite{typescript_json_schema}) minimizes some of these issues, but
future work should abstract our use of JSON Schema to other JSON DSL metalanguages, like TypeSchema~\cite{typeschema}.

\jsong{} supports a use case where expressivity and full DSL access is important, however, other usages might be well served by something form-like (\ala{} Flex-ER\cite{lobo2020flex}). Yet, tools like Ivy~\cite{mcnutt2021integrated} show that such modalities can fruitfully coexist.
Future work should explore how projections can mingle with simple forms like chart choosers~\cite{grammel_survey_2013}.

Beyond our case studies, there are numerous~\cite{schemaStore, baazizi2021empirical} exciting applications for \jsong{}.
For instance, supporting the web application DSL Varv~\cite{Varv22Borowski} might allow a self-hosting view system that is modifiable by the end-user.
A \jsong{}-based JSON Schema builder could guide users to create schemas that would make the schema more useful for \jsong{} (\eg{} including descriptions).
Aiding such domains promises to make a wide variety of novel experiences not only possible, but cheap.

\section{Acknowledgements}

We thank our survey participants for sharing their experiences, our reviewers for their thoughtful critiques, and the IEEE VIS Doctoral Colloquium for their support of this work. We also thank Will Brackenbury, Nicolas Kruchten, and Brian Hempel for their helpful comments on this work.

\bibliographystyle{IEEEtran}
\bibliography{IEEEabrv,prong-bib}

\begin{thebibliography}{10}
\providecommand{\url}[1]{#1}
\csname url@samestyle\endcsname
\providecommand{\newblock}{\relax}
\providecommand{\bibinfo}[2]{#2}
\providecommand{\BIBentrySTDinterwordspacing}{\spaceskip=0pt\relax}
\providecommand{\BIBentryALTinterwordstretchfactor}{4}
\providecommand{\BIBentryALTinterwordspacing}{\spaceskip=\fontdimen2\font plus
\BIBentryALTinterwordstretchfactor\fontdimen3\font minus
  \fontdimen4\font\relax}
\providecommand{\BIBforeignlanguage}[2]{{%
\expandafter\ifx\csname l@#1\endcsname\relax
\typeout{** WARNING: IEEEtran.bst: No hyphenation pattern has been}%
\typeout{** loaded for the language `#1'. Using the pattern for}%
\typeout{** the default language instead.}%
\else
\language=\csname l@#1\endcsname
\fi
#2}}
\providecommand{\BIBdecl}{\relax}
\BIBdecl

\bibitem{Fowler08Projectional}
\BIBentryALTinterwordspacing
M.~Fowler, ``Projectional editing,'' 2008. [Online]. Available:
  \url{https://martinfowler.com/bliki/ProjectionalEditing.html}
\BIBentrySTDinterwordspacing

\bibitem{horowitz2023live}
J.~Horowitz and J.~Heer, ``{Live, Rich, and Composable: Qualities for
  Programming Beyond Static Text},'' in \emph{PLATEAU Workshop}, 2023.

\bibitem{sns-deuce}
\BIBentryALTinterwordspacing
B.~Hempel, J.~Lubin, G.~Lu, and R.~Chugh, ``{Deuce: A Lightweight User
  Interface for Structured Editing},'' in \emph{International Conference on
  Software Engineering (ICSE)}.\hskip 1em plus 0.5em minus 0.4em\relax {ACM},
  2018, pp. 654--664. [Online]. Available:
  \url{https://doi.org/10.1145/3180155.3180165}
\BIBentrySTDinterwordspacing

\bibitem{hempel_sketch-n-sketch_2019}
\BIBentryALTinterwordspacing
B.~Hempel, J.~Lubin, and R.~Chugh, ``{Sketch-n-Sketch: Output-Directed
  Programming for SVG},'' in \emph{{Symposium on User Interface Software and
  Technology (UIST)}}.\hskip 1em plus 0.5em minus 0.4em\relax {ACM}, 2019, pp.
  281--292. [Online]. Available: \url{https://doi.org/10.1145/3332165.3347925}
\BIBentrySTDinterwordspacing

\bibitem{mongodb_inc_mongodb_nodate}
\BIBentryALTinterwordspacing
{MongoDB Inc.}, ``{MongoDB},'' 2023. [Online]. Available:
  \url{https://www.mongodb.com}
\BIBentrySTDinterwordspacing

\bibitem{mcnutt2022noGrammar}
\BIBentryALTinterwordspacing
A.~McNutt, ``{No Grammar to Rule Them All: A Survey of JSON-style DSLs for
  Visualization},'' \emph{{IEEE Transactions on Visualization and Computer
  Graphics}}, vol.~29, no.~1, pp. 160--170, 2023. [Online]. Available:
  \url{https://doi.org/10.1109/TVCG.2022.3209460}
\BIBentrySTDinterwordspacing

\bibitem{compton2015tracery}
\BIBentryALTinterwordspacing
K.~Compton, B.~A. Kybartas, and M.~Mateas, ``{Tracery: An Author-Focused
  Generative Text Tool},'' in \emph{{International Conference on Interactive
  Digital Storytelling {ICIDS}}}, vol. 9445.\hskip 1em plus 0.5em minus
  0.4em\relax Springer, 2015, pp. 154--161. [Online]. Available:
  \url{https://doi.org/10.1007/978-3-319-27036-4\_14}
\BIBentrySTDinterwordspacing

\bibitem{Varv22Borowski}
\BIBentryALTinterwordspacing
M.~Borowski, L.~Murray, R.~Bagge, J.~B. Kristensen, A.~Satyanarayan, and C.~N.
  Klokmose, ``{Varv: Reprogrammable Interactive Software As a Declarative Data
  Structure},'' in \emph{{Conference on Human Factors in Computing Systems
  (CHI)}}.\hskip 1em plus 0.5em minus 0.4em\relax {ACM}, 2022, pp.
  492:1--492:20. [Online]. Available:
  \url{https://doi.org/10.1145/3491102.3502064}
\BIBentrySTDinterwordspacing

\bibitem{payne2021danceon}
W.~C. Payne, Y.~Bergner, M.~E. West, C.~Charp, R.~B.~B. Shapiro, D.~A. Szafir,
  E.~V. Taylor, and K.~DesPortes, ``danceon: Culturally responsive creative
  computing,'' in \emph{{Conference on Human Factors in Computing Systems
  (CHI)}}.\hskip 1em plus 0.5em minus 0.4em\relax {ACM}, 2021, pp. 1--16.

\bibitem{pezoa_foundations_2016}
\BIBentryALTinterwordspacing
F.~Pezoa, J.~L. Reutter, F.~Su{\'{a}}rez, M.~Ugarte, and D.~Vrgoc,
  ``{Foundations of JSON Schema},'' in \emph{{International Conference on World
  Wide Web WWW}}.\hskip 1em plus 0.5em minus 0.4em\relax {ACM}, 2016, pp.
  263--273. [Online]. Available: \url{https://doi.org/10.1145/2872427.2883029}
\BIBentrySTDinterwordspacing

\bibitem{jakubovic2023Technical}
\BIBentryALTinterwordspacing
J.~Jakubovic, J.~Edwards, and T.~Petricek, ``{Technical Dimensions of
  Programming Systems},'' in \emph{{The Art, Science, and Engineering of
  Programming}}, vol.~7, no.~3, 2023. [Online]. Available:
  \url{https://doi.org/10.22152/programming-journal.org/2023/7/13}
\BIBentrySTDinterwordspacing

\bibitem{green1989cognitive}
T.~R. Green, ``{Cognitive Dimensions of Notations},'' \emph{People and
  computers V}, pp. 443--460, 1989.

\bibitem{interlispHistory}
\BIBentryALTinterwordspacing
W.~Teitelman, ``{History of Interlisp},'' in \emph{Celebrating the 50th
  Anniversary of Lisp}, ser. LISP50.\hskip 1em plus 0.5em minus 0.4em\relax
  ACM, 2008. [Online]. Available: \url{https://doi.org/10.1145/1529966.1529971}
\BIBentrySTDinterwordspacing

\bibitem{cornellProgramSynthesizer}
\BIBentryALTinterwordspacing
T.~Teitelbaum and T.~Reps, ``{The Cornell Program Synthesizer: A
  Syntax-Directed Programming Environment},'' \emph{Communications of the ACM},
  vol.~24, no.~9, p. 563–573, sep 1981. [Online]. Available:
  \url{https://doi.org/10.1145/358746.358755}
\BIBentrySTDinterwordspacing

\bibitem{resnick2009scratch}
\BIBentryALTinterwordspacing
M.~Resnick, J.~Maloney, A.~Monroy-Hern{\'a}ndez, N.~Rusk, E.~Eastmond,
  K.~Brennan, A.~Millner, E.~Rosenbaum, J.~Silver, B.~Silverman \emph{et~al.},
  ``{Scratch: Programming for All},'' \emph{Communications of the ACM},
  vol.~52, no.~11, pp. 60--67, 2009. [Online]. Available:
  \url{https://doi.org/10.1145/1592761.1592779}
\BIBentrySTDinterwordspacing

\bibitem{pech2013jetbrains}
\BIBentryALTinterwordspacing
V.~Pech, A.~Shatalin, and M.~Voelter, ``Jetbrains {MPS} as a tool for extending
  java,'' in \emph{Proceedings of the International Conference on Principles
  and Practices of Programming on the Java Platform: Virtual Machines,
  Languages, and Tools}.\hskip 1em plus 0.5em minus 0.4em\relax {ACM}, 2013,
  pp. 165--168. [Online]. Available:
  \url{https://doi.org/10.1145/2500828.2500846}
\BIBentrySTDinterwordspacing

\bibitem{tylr}
\BIBentryALTinterwordspacing
D.~Moon, A.~Blinn, and C.~Omar, ``Tylr: A tiny tile-based structure editor,''
  in \emph{SIGPLAN International Workshop on Type-Driven Development}, ser.
  TyDe 2022.\hskip 1em plus 0.5em minus 0.4em\relax {ACM}, 2022, pp. 28--37.
  [Online]. Available: \url{https://doi.org/10.1145/3546196.3550164}
\BIBentrySTDinterwordspacing

\bibitem{voinov2022forest}
\BIBentryALTinterwordspacing
P.~Voinov, M.~Rigger, and Z.~Su, ``Forest: Structural code editing with
  multiple cursors,'' in \emph{SIGPLAN International Symposium on New Ideas,
  New Paradigms, and Reflections on Programming and Software Onward!}\hskip 1em
  plus 0.5em minus 0.4em\relax {ACM}, 2022, pp. 137--152. [Online]. Available:
  \url{https://doi.org/10.1145/3563835.3567663}
\BIBentrySTDinterwordspacing

\bibitem{Beckmann23Sand}
\BIBentryALTinterwordspacing
T.~Beckmann, P.~Rein, S.~Ramson, J.~Bergsiek, and R.~Hirschfeld, ``{Structured
  Editing for All: Deriving Usable Structured Editors From Grammars},'' in
  \emph{{Conference on Human Factors in Computing Systems (CHI)}}.\hskip 1em
  plus 0.5em minus 0.4em\relax {ACM}, 2023, pp. 595:1--595:16. [Online].
  Available: \url{https://doi.org/10.1145/3544548.3580785}
\BIBentrySTDinterwordspacing

\bibitem{lerner2020projection}
\BIBentryALTinterwordspacing
S.~Lerner, ``{Projection Boxes: On-the-fly Reconfigurable Visualization for
  Live Programming},'' in \emph{{Conference on Human Factors in Computing
  Systems (CHI)}}.\hskip 1em plus 0.5em minus 0.4em\relax {ACM}, 2020, pp.
  1--7. [Online]. Available: \url{https://doi.org/10.1145/3313831.3376494}
\BIBentrySTDinterwordspacing

\bibitem{hoffswell_augmenting_2018}
\BIBentryALTinterwordspacing
J.~Hoffswell, A.~Satyanarayan, and J.~Heer, ``Augmenting code with in situ
  visualizations to aid program understanding,'' in \emph{{Conference on Human
  Factors in Computing Systems (CHI)}}.\hskip 1em plus 0.5em minus 0.4em\relax
  {ACM}, 2018, p. 1–12. [Online]. Available:
  \url{https://doi.org/10.1145/3173574.3174106}
\BIBentrySTDinterwordspacing

\bibitem{gobert2022latex}
\BIBentryALTinterwordspacing
C.~Gobert and M.~Beaudouin-Lafon, ``{i-LaTeX: Manipulating Transitional
  Representations between LaTeX Code and Generated Documents},'' in
  \emph{Proceedings of the 2022 CHI Conference on Human Factors in Computing
  Systems}, ser. CHI '22.\hskip 1em plus 0.5em minus 0.4em\relax {ACM}, 2022.
  [Online]. Available: \url{https://doi.org/10.1145/3491102.3517494}
\BIBentrySTDinterwordspacing

\bibitem{awesomeStructureEditors}
\BIBentryALTinterwordspacing
Y.~Chuchem, ``{awesome-structure-editors},'' 2022. [Online]. Available:
  \url{https://github.com/yairchu/awesome-structure-editors}
\BIBentrySTDinterwordspacing

\bibitem{ko2005citrus}
\BIBentryALTinterwordspacing
A.~J. Ko and B.~A. Myers, ``{Citrus: a Language and Toolkit for Simplifying the
  Creation of Structured Editors for Code and Data},'' in \emph{{Symposium on
  User Interface Software and Technology (UIST)}}.\hskip 1em plus 0.5em minus
  0.4em\relax {ACM}, 2005, pp. 3--12. [Online]. Available:
  \url{https://doi.org/10.1145/1095034.1095037}
\BIBentrySTDinterwordspacing

\bibitem{ko2006barista}
\BIBentryALTinterwordspacing
------, ``{Barista: An Implementation Framework for Enabling New Tools,
  Interaction Techniques and Views in Code Editors},'' in \emph{{Conference on
  Human Factors in Computing Systems (CHI)}}.\hskip 1em plus 0.5em minus
  0.4em\relax {ACM}, 2006, pp. 387--396. [Online]. Available:
  \url{https://doi.org/10.1145/1124772.1124831}
\BIBentrySTDinterwordspacing

\bibitem{omar2012active}
\BIBentryALTinterwordspacing
C.~Omar, Y.~Yoon, T.~D. LaToza, and B.~A. Myers, ``{Active Code Completion},''
  in \emph{International Conference on Software Engineering (ICSE)}.\hskip 1em
  plus 0.5em minus 0.4em\relax {IEEE}, 2012, pp. 859--869. [Online]. Available:
  \url{https://doi.org/10.1109/ICSE.2012.6227133}
\BIBentrySTDinterwordspacing

\bibitem{andersen2020adding}
\BIBentryALTinterwordspacing
L.~Andersen, M.~Ballantyne, and M.~Felleisen, ``{Adding Interactive Visual
  Syntax to Textual Code},'' \emph{International Conference on Object Oriented
  Programming Systems Languages {\&} Applications {OOPSLA}}, vol.~4, pp.
  222:1--222:28, 2020. [Online]. Available:
  \url{https://doi.org/10.1145/3428290}
\BIBentrySTDinterwordspacing

\bibitem{ferdowsifard2020small}
\BIBentryALTinterwordspacing
K.~Ferdowsifard, A.~Ordookhanians, H.~Peleg, S.~Lerner, and N.~Polikarpova,
  ``{Small-Step Live Programming by Example},'' in \emph{{Symposium on User
  Interface Software and Technology (UIST)}}.\hskip 1em plus 0.5em minus
  0.4em\relax {ACM}, 2020, pp. 614--626. [Online]. Available:
  \url{https://doi.org/10.1145/3379337.3415869}
\BIBentrySTDinterwordspacing

\bibitem{omar2021filling}
\BIBentryALTinterwordspacing
C.~Omar, D.~Moon, A.~Blinn, I.~Voysey, N.~Collins, and R.~Chugh, ``Filling
  typed holes with live guis,'' in \emph{SIGPLAN International Conference on
  Programming Language Design and Implementation (PLDI)}.\hskip 1em plus 0.5em
  minus 0.4em\relax {ACM}, 2021, pp. 511--525. [Online]. Available:
  \url{https://doi.org/10.1145/3453483.3454059}
\BIBentrySTDinterwordspacing

\bibitem{Zong21Lyra2}
\BIBentryALTinterwordspacing
J.~Zong, D.~Barnwal, R.~Neogy, and A.~Satyanarayan, ``{Lyra 2: Designing
  Interactive Visualizations by Demonstration},'' \emph{{{IEEE Transactions on
  Visualization and Computer Graphics}}}, vol.~27, no.~2, pp. 304--314, 2021.
  [Online]. Available: \url{https://doi.org/10.1109/TVCG.2020.3030367}
\BIBentrySTDinterwordspacing

\bibitem{dataworld_chart_nodate}
\BIBentryALTinterwordspacing
data.world, ``{Chart Builder}.'' [Online]. Available:
  \url{https://github.com/datadotworld/chart-builder}
\BIBentrySTDinterwordspacing

\bibitem{hoffswell2020techniques}
\BIBentryALTinterwordspacing
J.~Hoffswell, W.~Li, and Z.~Liu, ``{Techniques for Flexible Responsive
  Visualization Design},'' in \emph{{Conference on Human Factors in Computing
  Systems (CHI)}}.\hskip 1em plus 0.5em minus 0.4em\relax {ACM}, 2020, pp.
  1--13. [Online]. Available: \url{https://doi.org/10.1145/3313831.3376777}
\BIBentrySTDinterwordspacing

\bibitem{jsoneditor}
\BIBentryALTinterwordspacing
J.~de~Jong, ``{jsoneditor},'' 2021. [Online]. Available:
  \url{https://github.com/josdejong/jsoneditor}
\BIBentrySTDinterwordspacing

\bibitem{jetPenner}
\BIBentryALTinterwordspacing
C.~Penner, ``{jet},'' 2021. [Online]. Available:
  \url{https://github.com/ChrisPenner/jet/}
\BIBentrySTDinterwordspacing

\bibitem{noauthor_editoride_nodate}
\BIBentryALTinterwordspacing
Vega, ``{Editor/IDE for Vega and Vega-Lite}.'' [Online]. Available:
  \url{https://vega.github.io/editor/}
\BIBentrySTDinterwordspacing

\bibitem{LittRunTimeSparklines}
\BIBentryALTinterwordspacing
G.~Litt, ``{Runtime Visualization for Model-View-Update GUIs},'' 2020.
  [Online]. Available:
  \url{https://www.geoffreylitt.com/resources/todomvc-vis.pdf}
\BIBentrySTDinterwordspacing

\bibitem{Travers21Vaguely}
\BIBentryALTinterwordspacing
M.~Travers, ``Vaguely,'' 2021. [Online]. Available:
  \url{https://github.com/mtravers/vaguely}
\BIBentrySTDinterwordspacing

\bibitem{chavarriaga2023approach}
\BIBentryALTinterwordspacing
E.~Chavarriaga, F.~Jurado, and F.~D. Rodr{\'{\i}}guez, ``An approach to build
  json-based domain specific languages solutions for web applications,''
  \emph{Journal of Computer Languages}, vol.~75, pp. 1--18, 2023. [Online].
  Available: \url{https://doi.org/10.1016/j.cola.2023.101203}
\BIBentrySTDinterwordspacing

\bibitem{petitpierre2011bottom}
\BIBentryALTinterwordspacing
C.~Petitpierre, ``Bottom up creation of a {DSL} using templates and {JSON},''
  in \emph{SPLASH'11 Workshops - Compilation Proceedings of the Co-Located
  Workshops: DSM'11, TMC'11, AGERE!'11, AOOPES'11, NEAT'11, and VMIL'11}.\hskip
  1em plus 0.5em minus 0.4em\relax {ACM}, 2011, pp. 47--52. [Online].
  Available: \url{https://doi.org/10.1145/2095050.2095059}
\BIBentrySTDinterwordspacing

\bibitem{vscodeJSONLSP}
\BIBentryALTinterwordspacing
Microsoft, ``{vscode-json-languageservice},'' 2021. [Online]. Available:
  \url{https://github.com/microsoft/vscode-json-languageservice}
\BIBentrySTDinterwordspacing

\bibitem{deneb}
\BIBentryALTinterwordspacing
D.~Marsh-Patrick, ``{deneb}.'' [Online]. Available:
  \url{https://deneb-viz.github.io/visual-editor}
\BIBentrySTDinterwordspacing

\bibitem{mongod_compass}
\BIBentryALTinterwordspacing
{MongoDB Inc.}, ``{Compass. The GUI for MongoDB.}'' 2023. [Online]. Available:
  \url{https://www.mongodb.com/products/compass}
\BIBentrySTDinterwordspacing

\bibitem{traceryEditor}
\BIBentryALTinterwordspacing
K.~Compton, ``{Tracery Editor},'' 2021. [Online]. Available:
  \url{http://tracery.io/editor/}
\BIBentrySTDinterwordspacing

\bibitem{whitworth_polite_2005}
\BIBentryALTinterwordspacing
B.~Whitworth, ``{Polite Computing},'' \emph{Behaviour \& Information
  Technology}, vol.~24, no.~5, pp. 353--363, 2005. [Online]. Available:
  \url{https://doi.org/10.1080/01449290512331333700}
\BIBentrySTDinterwordspacing

\bibitem{chart_studio_plotly}
\BIBentryALTinterwordspacing
{plotly}, ``{Chart Studio},'' 2023. [Online]. Available:
  \url{https://chart-studio.plotly.com/create/#/}
\BIBentrySTDinterwordspacing

\bibitem{hoffswell_visual_2016}
\BIBentryALTinterwordspacing
J.~Hoffswell, A.~Satyanarayan, and J.~Heer, ``{Visual Debugging Techniques for
  Reactive Data Visualization},'' in \emph{Computer {Graphics} {Forum}},
  vol.~35, no.~3.\hskip 1em plus 0.5em minus 0.4em\relax Wiley Online Library,
  2016, pp. 271--280. [Online]. Available:
  \url{https://doi.org/10.1111/cgf.12903}
\BIBentrySTDinterwordspacing

\bibitem{mcnutt_surfacing_2020}
\BIBentryALTinterwordspacing
A.~McNutt, G.~Kindlmann, and M.~Correll, ``Surfacing visualization mirages,''
  \emph{{Conference on Human Factors in Computing Systems (CHI)}}, pp. 1--16,
  2020. [Online]. Available: \url{https://doi.org/10.1145/3313831.3376420}
\BIBentrySTDinterwordspacing

\bibitem{hopkins_visualint_2020}
\BIBentryALTinterwordspacing
A.~K. Hopkins, M.~Correll, and A.~Satyanarayan, ``{VisuaLint: Sketchy In Situ
  Annotations of Chart Construction Errors},'' vol.~39, no.~3, 2020, pp.
  219--228. [Online]. Available: \url{https://doi.org/10.1111/cgf.13975}
\BIBentrySTDinterwordspacing

\bibitem{chen2021vizlinter}
\BIBentryALTinterwordspacing
Q.~Chen, F.~Sun, X.~Xu, Z.~Chen, J.~Wang, and N.~Cao, ``{VizLinter: A Linter
  and Fixer Framework for Data Visualization},'' \emph{{IEEE Transactions on
  Visualization and Computer Graphics}}, vol.~28, no.~1, pp. 206--216, 2022.
  [Online]. Available: \url{https://doi.org/10.1109/TVCG.2021.3114804}
\BIBentrySTDinterwordspacing

\bibitem{mcnutt_linting_2018}
A.~McNutt and G.~Kindlmann, ``{Linting for Visualization: Towards a Practical
  Automated Visualization Guidance System},'' in \emph{{VisGuides: 2nd Workshop
  on the Creation, Curation, Critique and Conditioning of Principles and
  Guidelines in Visualization}}, 2018.

\bibitem{mcnutt2021integrated}
A.~McNutt and R.~Chugh, ``{Integrated Visualization Editing via Parameterized
  Declarative Templates},'' in \emph{{Conference on Human Factors in Computing
  Systems (CHI)}}.\hskip 1em plus 0.5em minus 0.4em\relax {ACM}, 2021, pp.
  1--14.

\bibitem{guzdial2021integrating}
\BIBentryALTinterwordspacing
M.~Guzdial and T.~Shreiner, ``Integrating computing through task-specific
  programming for disciplinary relevance: Considerations and examples,''
  \emph{Computational Thinking in Compulsory Education}, 2021. [Online].
  Available: \url{http://dx.doi.org/10.4324/9781003102991-10}
\BIBentrySTDinterwordspacing

\bibitem{brandt_two_2009}
\BIBentryALTinterwordspacing
J.~Brandt, P.~J. Guo, J.~Lewenstein, M.~Dontcheva, and S.~R. Klemmer, ``Two
  {Studies} of {Opportunistic} {Programming}: {Interleaving} {Web} {Foraging},
  {Learning}, and {Writing} {Code},'' in \emph{{Conference on Human Factors in
  Computing Systems (CHI)}}.\hskip 1em plus 0.5em minus 0.4em\relax {ACM},
  2009, pp. 1589--1598. [Online]. Available:
  \url{https://doi.org/10.1145/1518701.1518944}
\BIBentrySTDinterwordspacing

\bibitem{grammel_survey_2013}
\BIBentryALTinterwordspacing
L.~Grammel, C.~Bennett, M.~Tory, and M.-A.~D. Storey, ``{A Survey of
  Visualization Construction User Interfaces},'' in \emph{{EuroVis} ({Short}
  {Papers})}, 2013. [Online]. Available:
  \url{https://doi.org/10.2312/PE.EuroVisShort.EuroVisShort2013.019-023}
\BIBentrySTDinterwordspacing

\bibitem{codemirror}
\BIBentryALTinterwordspacing
M.~Haverbeke, ``Code mirror: Extensible code editor,'' 2023. [Online].
  Available: \url{https://codemirror.net/}
\BIBentrySTDinterwordspacing

\bibitem{ajv}
\BIBentryALTinterwordspacing
{AJV}, ``{Ajv JSON schema validator},'' 2023. [Online]. Available:
  \url{https://ajv.js.org/}
\BIBentrySTDinterwordspacing

\bibitem{schemaStore}
\BIBentryALTinterwordspacing
SchemaStore, ``Json schema store.'' [Online]. Available:
  \url{https://github.com/schemastore/schemastore/}
\BIBentrySTDinterwordspacing

\bibitem{oney2012codelets}
\BIBentryALTinterwordspacing
S.~Oney and J.~Brandt, ``Codelets: Linking interactive documentation and
  example code in the editor,'' in \emph{{Conference on Human Factors in
  Computing Systems (CHI)}}, ser. CHI '12.\hskip 1em plus 0.5em minus
  0.4em\relax {ACM}, 2012, pp. 2697--2706. [Online]. Available:
  \url{https://doi.org/10.1145/2207676.2208664}
\BIBentrySTDinterwordspacing

\bibitem{rong2016codemend}
\BIBentryALTinterwordspacing
X.~Rong, S.~Yan, S.~Oney, M.~Dontcheva, and E.~Adar, ``{Codemend: Assisting
  Interactive Programming with Bimodal Embedding},'' in \emph{{Symposium on
  User Interface Software and Technology (UIST)}}.\hskip 1em plus 0.5em minus
  0.4em\relax {ACM}, 2016, pp. 247--258. [Online]. Available:
  \url{https://doi.org/10.1145/2984511.2984544}
\BIBentrySTDinterwordspacing

\bibitem{goffin2016exploratory}
\BIBentryALTinterwordspacing
P.~Goffin, J.~Boy, W.~Willett, and P.~Isenberg, ``{An Exploratory Study of
  Word-scale Graphics in Data-rich Text Documents},'' \emph{{{IEEE Transactions
  on Visualization and Computer Graphics}}}, vol.~23, no.~10, pp. 2275--2287,
  2017. [Online]. Available: \url{https://doi.org/10.1109/TVCG.2016.2618797}
\BIBentrySTDinterwordspacing

\bibitem{naimipour2020engaging}
\BIBentryALTinterwordspacing
B.~Naimipour, M.~Guzdial, and T.~Shreiner, ``{Engaging Pre-service Teachers in
  Front-end Design: Developing Technology for a Social Studies Classroom},'' in
  \emph{{Frontiers in Education Conference}}.\hskip 1em plus 0.5em minus
  0.4em\relax IEEE, 2020, pp. 1--9. [Online]. Available:
  \url{https://doi.org/10.1109/FIE44824.2020.9273908}
\BIBentrySTDinterwordspacing

\bibitem{nelson2021creating}
\BIBentryALTinterwordspacing
T.~Nelson-Fromm, ``{Creating Better Teaching Tools Through Examining Teachers'
  Understanding of Data Representations},'' in \emph{Symposium on Visual
  Languages and Human-Centric Computing (VL/HCC)}.\hskip 1em plus 0.5em minus
  0.4em\relax IEEE, 2021, pp. 1--2. [Online]. Available:
  \url{https://doi.org/10.1109/VL/HCC51201.2021.9576206}
\BIBentrySTDinterwordspacing

\bibitem{wongsuphasawat_voyager_2017}
\BIBentryALTinterwordspacing
K.~Wongsuphasawat, Z.~Qu, D.~Moritz, R.~Chang, F.~Ouk, A.~Anand, J.~Mackinlay,
  B.~Howe, and J.~Heer, ``{Voyager 2: Augmenting Visual Analysis with Partial
  View Specifications},'' in \emph{{Conference on Human Factors in Computing
  Systems (CHI)}}.\hskip 1em plus 0.5em minus 0.4em\relax {ACM}, 2017, pp.
  2648--2659. [Online]. Available:
  \url{https://doi.org/10.1145/3025453.3025768}
\BIBentrySTDinterwordspacing

\bibitem{fowler2010domain}
M.~Fowler, \emph{Domain-specific languages}.\hskip 1em plus 0.5em minus
  0.4em\relax Pearson Education, 2010.

\bibitem{barke2022grounded}
\BIBentryALTinterwordspacing
S.~Barke, M.~B. James, and N.~Polikarpova, ``Grounded copilot: How programmers
  interact with code-generating models,'' \emph{International Conference on
  Object Oriented Programming Systems Languages {\&} Applications {OOPSLA}},
  vol.~7, no. {OOPSLA1}, pp. 85--111, 2023. [Online]. Available:
  \url{https://doi.org/10.1145/3586030}
\BIBentrySTDinterwordspacing

\bibitem{blinn2022integrative}
\BIBentryALTinterwordspacing
A.~Blinn, D.~Moon, E.~Griffis, and C.~Omar, ``{An Integrative Human-Centered
  Architecture for Interactive Programming Assistants},'' in \emph{Symposium on
  Visual Languages and Human-Centric Computing (VL/HCC)}.\hskip 1em plus 0.5em
  minus 0.4em\relax IEEE, 2022, pp. 1--5. [Online]. Available:
  \url{https://doi.org/10.1109/VL/HCC53370.2022.9833110}
\BIBentrySTDinterwordspacing

\bibitem{comptonTraceryMiddle}
K.~Compton, ``tweet,'' May 2022,
  \url{https://twitter.com/GalaxyKate/status/1528109593187975168?s=20}
  content:\\ Thanks to my Patreon supporters, today you paid for 30 Chicago
  middle school girls to learn Tracery, I guess.

\bibitem{beaudouin2004designing}
\BIBentryALTinterwordspacing
M.~Beaudouin{-}Lafon, ``{Designing Interaction, Not Interfaces},'' in
  \emph{{Workshop Conference on Advanced Visual Interfaces (AVI)}}.\hskip 1em
  plus 0.5em minus 0.4em\relax {ACM}, 2004. [Online]. Available:
  \url{https://doi.org/10.1145/989863.989865}
\BIBentrySTDinterwordspacing

\bibitem{jsoncParser}
\BIBentryALTinterwordspacing
Microsoft, ``{node-jsonc-parser}.'' [Online]. Available:
  \url{https://github.com/microsoft/node-jsonc-parser}
\BIBentrySTDinterwordspacing

\bibitem{gathani2020debugging}
\BIBentryALTinterwordspacing
S.~Gathani, P.~Lim, and L.~Battle, ``Debugging database queries: A survey of
  tools, techniques, and users,'' in \emph{{Conference on Human Factors in
  Computing Systems (CHI)}}, ser. CHI '20.\hskip 1em plus 0.5em minus
  0.4em\relax {ACM}, 2020, pp. 1--16. [Online]. Available:
  \url{https://doi.org/10.1145/3313831.3376485}
\BIBentrySTDinterwordspacing

\bibitem{shen2022galvis}
\BIBentryALTinterwordspacing
L.~Shen, E.~Shen, Z.~Tai, Y.~Wang, Y.~Luo, and J.~Wang, ``{GALVIS:
  Visualization Construction through Example-Powered Declarative
  Programming},'' in \emph{International Conference on Information \& Knowledge
  Management}.\hskip 1em plus 0.5em minus 0.4em\relax {ACM}, 2022, pp.
  4975--4979. [Online]. Available:
  \url{https://doi.org/10.1145/3511808.3557159}
\BIBentrySTDinterwordspacing

\bibitem{berger2016efficiency}
\BIBentryALTinterwordspacing
T.~Berger, M.~V\"{o}lter, H.~P. Jensen, T.~Dangprasert, and J.~Siegmund,
  ``{Efficiency of Projectional Editing: A Controlled Experiment},'' in
  \emph{SIGSOFT International Symposium on Foundations of Software
  Engineering}.\hskip 1em plus 0.5em minus 0.4em\relax {ACM}, 2016, pp.
  763--774. [Online]. Available: \url{https://doi.org/10.1145/2950290.2950315}
\BIBentrySTDinterwordspacing

\bibitem{chodarev2022experimental}
\BIBentryALTinterwordspacing
S.~Chodarev, M.~Sul{\'\i}r, J.~Porub{\"a}n, and M.~Kop{\v{c}}{\'a}kov{\'a},
  ``Experimental comparison of editor types for domain-specific languages,''
  \emph{Applied Sciences}, vol.~12, no.~19, 2022. [Online]. Available:
  \url{https://www.mdpi.com/2076-3417/12/19/9893}
\BIBentrySTDinterwordspacing

\bibitem{satyanarayan_critical_2020}
\BIBentryALTinterwordspacing
A.~Satyanarayan, B.~Lee, D.~Ren, J.~Heer, J.~T. Stasko, J.~Thompson,
  M.~Brehmer, and Z.~Liu, ``{Critical Reflections on Visualization Authoring
  Systems},'' \emph{{IEEE Transactions on Visualization and Computer
  Graphics}}, vol.~26, no.~1, pp. 461--471, 2020. [Online]. Available:
  \url{https://doi.org/10.1109/TVCG.2019.2934281}
\BIBentrySTDinterwordspacing

\bibitem{sulir2018visual}
M.~Sul{\'\i}r, M.~Ba{\v{c}}{\'\i}kov{\'a}, S.~Chodarev, and J.~Porub{\"a}n,
  ``Visual augmentation of source code editors: A systematic mapping study,''
  \emph{Journal of Visual Languages \& Computing}, vol.~49, pp. 46--59, 2018.

\bibitem{JSONSchemaIssue}
\BIBentryALTinterwordspacing
Stackoverflow, ``{JSON schema : ``allof'' with ``additionalProperties''},''
  2015. [Online]. Available:
  \url{https://stackoverflow.com/questions/22689900/json-schema-allof-with-additionalproperties}
\BIBentrySTDinterwordspacing

\bibitem{baazizi2021empirical}
\BIBentryALTinterwordspacing
M.~A. Baazizi, D.~Colazzo, G.~Ghelli, C.~Sartiani, and S.~Scherzinger, ``{An
  Empirical Study on the "Usage of Not" in Real-world JSON Schema Documents},''
  in \emph{International Conference on Conceptual Modeling}, vol. 13011.\hskip
  1em plus 0.5em minus 0.4em\relax Springer, 2021, pp. 102--112. [Online].
  Available: \url{https://doi.org/10.1007/978-3-030-89022-3\_9}
\BIBentrySTDinterwordspacing

\bibitem{typescript_json_schema}
\BIBentryALTinterwordspacing
YousefED, ``{typescript-json-schema}.'' [Online]. Available:
  \url{https://github.com/YousefED/typescript-json-schema}
\BIBentrySTDinterwordspacing

\bibitem{typeschema}
\BIBentryALTinterwordspacing
apioo, ``{TypeSchema}.'' [Online]. Available: \url{https://typeschema.org/}
\BIBentrySTDinterwordspacing

\bibitem{lobo2020flex}
\BIBentryALTinterwordspacing
M.~J. Lobo, C.~Hurter, and P.~Irani, ``Flex-er: A platform to evaluate
  interaction techniques for immersive visualizations,'' \emph{Human-Computer
  Interaction}, vol.~4, no. ISS, nov 2020. [Online]. Available:
  \url{https://doi.org/10.1145/3427323}
\BIBentrySTDinterwordspacing

\end{thebibliography}

\end{document}